\pdfoutput=1
\documentclass[11pt,twoside,a4paper,cmspaper,final,collab]{cms-tdr}
\def\svnVersion{74d5a23}\def\svnDate{2025/10/01}\def\cmsCernNoTag{CERN-EP-2024-328}\def\cmsCernDate{\today}\def\cmsMessage{Published in Physical Review Letters as \href{http://dx.doi.org/10.1103/6ywq-gm61}{\doi{10.1103/6ywq-gm61}.}}
\begin{document}\cmsNoteHeader{HIN-24-002}

\newlength\cmsTabSkip\setlength{\cmsTabSkip}{1ex}
\newcommand{\noff}    {\ensuremath{N_\text{trk}^\text{offline}}\xspace}
\newcommand {\AonA}  {\ensuremath{\text{AA}}\xspace}
\newcommand {\PbPb}  {\ensuremath{\text{PbPb}}\xspace}
\newcommand {\pPb}  {\ensuremath{\Pp\text{Pb}}\xspace}
\newcommand {\pp}    {\ensuremath{\Pp\Pp}\xspace}
\newcommand {\pA}    {\ensuremath{\Pp\text{A}}\xspace}
\newcommand{\ncorr}    {\ensuremath{N_\text{trk}^\text{corrected}}\xspace}
\newcommand{\ncorravg} {\ensuremath{\langle \ncorr \rangle}\xspace}
\newcommand{\Pzs}{\ensuremath{P_{z,\mathrm{s}2}}\xspace}
\newcommand{\Pz}{\ensuremath{P_{z}}\xspace}
\newcommand{\phii}{\ensuremath{\phi_{i}}\xspace}
\newcommand{\omegai}{\ensuremath{\omega_{i}}\xspace}
\newcommand{\Qnx} {\ensuremath{Q_{n,x}}\xspace}
\newcommand{\Qny} {\ensuremath{Q_{n,y}}\xspace}
\newcommand{\Psix}[1] {\ensuremath{\Psi_{#1}}\xspace}
\ifthenelse{\boolean{cms@external}}{\providecommand{\cmsleft}{upper\xspace}}{\providecommand{\cmsleft}{left\xspace}}
\ifthenelse{\boolean{cms@external}}{\providecommand{\cmsright}{lower\xspace}}{\providecommand{\cmsright}{right\xspace}}
\ifthenelse{\boolean{cms@external}}{\providecommand{\cmsLeft}{Upper\xspace}}{\providecommand{\cmsLeft}{Left\xspace}}
\ifthenelse{\boolean{cms@external}}{\providecommand{\cmsRight}{Lower\xspace}}{\providecommand{\cmsRight}{Right\xspace}}

\ifthenelse{\boolean{cms@external}}{%
    \newcommand{\refsupplemental}{the supplemental material~\cite{this:supplemental}\xspace}
}{%
	\newcommand{\refsupplemental}{the supplemental material\nocite{placeholder}\xspace}
}

\title{
	Observation of \texorpdfstring{\PGL}{Lambda} hyperon local polarization in \texorpdfstring{\pPb}{pPb} collisions at \texorpdfstring{$\sqrtsNN = 8.16\TeV$}{sqrtsNN = 8.16 TeV}}

\date{\today}

\abstract{
The polarization of the \PGL and \PAGL hyperons along the beam direction has been measured in proton-lead (\pPb) collisions at a center-of-mass energy per nucleon pair of 8.16\TeV.  The data were obtained with the CMS detector at the LHC and correspond to an integrated luminosity of $186.0\pm6.5\nbinv$. A significant azimuthal dependence of the hyperon polarization, characterized by the second-order Fourier sine coefficient \Pzs, is observed.  The \Pzs values decrease as a function of charged particle multiplicity, but increase with transverse momentum. A hydrodynamic model that describes the observed \Pzs values in nucleus-nucleus collisions by introducing vorticity effects does not reproduce either the sign or the magnitude of the \pPb results. These observations pose a challenge to the current theoretical implementation of spin polarization in heavy ion collisions and offer new insights into the origin of spin polarization in hadronic collisions at LHC energies.
}

\hypersetup{
pdfauthor={CMS Collaboration},
pdftitle={Observation of lambda hyperon local polarization in pPb collisions at sqrt(s[NN]) = 8.16 TeV},
pdfsubject={CMS},
pdfkeywords={CMS, heavy ion, hyperon polarization, small system}} 
\maketitle

Quantum chromodynamics (QCD) predicts that nuclear matter will undergo a phase transition to a deconfined partonic phase, the so-called quark-gluon plasma (QGP), 
at very high density and temperature~\cite{Shuryak:1977ut,Collins:1974ky}. 
This state of matter is generated in ultrarelativistic collisions of heavy nuclei and behaves like an almost ideal fluid~\cite{Braun-Munzinger:2015hba}. 
The nontrivial velocity and vorticity fields developed in the evolution of the QGP are predicted to induce polarization 
of produced particles via spin-orbit coupling~\cite{Liang:2004ph,Becattini:2007sr}, 
analogous to the Barnett effect~\cite{Barnett:1935wyv}, which is the magnetization of a material induced by mechanical rotation. 
In particular, the nonzero vorticity component resulting from the shear in the initial velocity distributions of the participants in off-center nucleus-nucleus (\AonA) 
collisions can generate a net particle polarization perpendicular to the plane spanned by the beam direction ($z$) and the impact parameter.
This polarization, referred to as the global polarization, has a direction along the orbital angular momentum of the initial system.
The observations and studies of the \PGL hyperon global polarization in heavy ion collisions 
at the Brookhaven National Laboratory Relativistic Heavy Ion Collider (RHIC)~\cite{STAR:2017ckg,STAR:2019erd} and the CERN LHC \cite{ALICE:2019onw} have opened new directions for the study of the QGP fluid and its spin dynamics. 

Polarization is also possible along the beam direction.  
In \AonA collisions, pressure gradients within the initially asymmetric geometry of the created QGP lead to azimuthally anisotropic particle emission 
relative to a global event-plane direction~\cite{Ollitrault:1992bk,Heinz:2013th,Gale:2013da}. 
This anisotropic collective flow is predicted to generate nonzero vorticity along the beam axis that, in turn, 
results in particle polarization along the beam direction (\Pz)~\cite{Voloshin:2017kqp,Becattini:2017gcx}. 
This ``local'' polarization~\cite{Gao:2012ix,Xia:2018tes} has a sinusoidal dependence on the azimuth of the hyperon with respect to the $n\mathrm{th}$-order event-plane angle \Psix{n}, determined by the direction of maximum particle density for each corresponding order.
No net polarization along the beam axis is expected after averaging over the full azimuth, because of the geometrical symmetry. 
The sinusoidal dependence of \Pz can be characterized by the Fourier harmonic $P_{z,\mathrm{s}n} = \langle \Pz\sin[n(\phi-\Psix{n})] \rangle$, 
where $\phi$ is the hyperon azimuthal emission angle.
The angular brackets indicate an average over all hyperons in all selected events. 
The \PGL hyperon \Pzs  value was first measured in gold-gold collisions at $\sqrtsNN = 200\GeV$~\cite{STAR:2019erd} and later in lead-lead (\PbPb) collisions at $\sqrtsNN = 5.02\TeV$~\cite{ALICE:2021pzu}. 
Recently, similar polarization has been observed with respect to the third-order event plane in ruthenium-ruthenium and zirconium-zirconium collisions at $\sqrtsNN = 200\GeV$~\cite{STAR:2023eck}.
Within the framework of hydrodynamic and transport models, it has been found that the inclusion of shear-induced polarization is required to capture the sign and magnitude 
of the experimental results~\cite{Becattini:2017gcx,Xia:2018tes,Fu:2021pok,Becattini:2021iol,Alzhrani:2022dpi,Florkowski:2021xvy}. 
However, the calculations are very sensitive to the implementation details of the shear contributions~\cite{Alzhrani:2022dpi,Florkowski:2021xvy}.

The observation of collective flow signals in events with high final-state particle multiplicity in proton-proton (\pp)~\cite{Khachatryan:2010gv,Aad:2015gqa,Khachatryan:2015lva,Khachatryan:2016txc,CMS:2020qul} and proton-lead (\pPb)~\cite{CMS:2012qk,alice:2012qe,Aad:2012gla,LHCb:2015coe,CMS:2018loe,CMS:2018duw,CMS:2022bmk,CMS:2024krd} collisions at the LHC, as well as at RHIC in proton-gold, deuteron-gold, and $^{3}{\mathrm{He}}$-gold collisions~\cite{PHENIX:2014fnc,PHENIX:2015idk,PHENIX:2016cfs,PHENIX:2017djs,PHENIX:2018lia,PHENIX:2021ubk,STAR:2015kak,STAR:2019zaf,STAR:2022pfn}, has raised the question of whether a fluidlike QGP is created in 
these lighter hadronic collision systems~\cite{Nagle:2018nvi, Dusling:2015gta}.
If that is the case, the azimuthal-dependent pressure gradients will lead to vorticity structures similar to those in large \AonA collisions. 
It is therefore of great interest to investigate whether such a vorticity is created in a system like the one formed in high-multiplicity \pPb collisions. 
Measuring \PGL hyperon polarization along the beam direction in \pPb interactions can also provide new constraints on the interpretation of the observed hyperon polarization in \AonA collisions.

This Letter presents the first measurement of \PGL and \PAGL hyperon polarization along the beam direction in \pPb collisions at $\sqrtsNN = 8.16\TeV$.
The \Pzs values for \PGL and \PAGL particles with $\abs{\eta}<2.4$ and transverse momentum $0.8<\pt<6.0\GeVc$ are reported  
as a function of \pt and the event charged particle multiplicity. Tabulated results are provided in the HEPData record for this analysis~\cite{hepdata}.

The central feature of the CMS apparatus is a superconducting solenoid of 6\unit{m}
internal diameter, providing a magnetic field of 3.8\unit{T}. Within the solenoid volume,
there are four subdetectors, including a silicon
pixel and strip tracker detector, a lead tungstate crystal electromagnetic calorimeter (ECAL) with around 4000 towers,
and a brass and scintillator hadron calorimeter (HCAL) with around 4300 towers, each composed of a
barrel and two endcap sections. Iron and quartz-fiber Cherenkov hadron forward (HF)
calorimeters cover the pseudorapidity range $2.9<\abs{\eta}<5.2$.
The silicon tracker measures charged particles within the range $\abs{\eta}<2.5$. For charged
particles with $1<\pt<10\GeVc$ and $\abs{\eta}<1.4$, the track resolutions are
typically 1.5\% in \pt and 25--90 (45--150)\mum in the transverse (longitudinal)
impact parameter \cite{CMS:2014pgm}. A detailed description of the Compact Muon Solenoid (CMS) detector,
together with a definition of the coordinate system used and the relevant kinematic variables,
can be found in Ref. \cite{Chatrchyan:2008zzk}.

The \pPb  data at $\sqrtsNN = 8.16\TeV$ used in this analysis were
collected by the CMS experiment in 2016, and correspond to an integrated luminosity of $186.0\pm6.5\nbinv$~\cite{CMS:2018fkg}. 
The beam energies are 6.5\TeV for the protons and
2.56\TeV per nucleon for the lead nuclei. The event reconstruction, event selections, and triggers are identical to
those described in Refs.~\cite{Khachatryan:2016txc,CMS:2017lrw,Sirunyan:2017uyl}. 
The minimum bias (MB) events are triggered by requiring energy
deposits above 1\GeV in at least one of the two hadronic forward (HF) calorimeters and
the presence of at least one track with $\pt>0.4\GeVc$ reconstructed using hits from only the pixel tracker. 
Dedicated high-multiplicity (HM) triggers are implemented to collect a large event sample featuring a high number of final-state particles. 
The total number of combined ECAL and HCAL towers having deposited energy above a threshold of 0.5\GeV in transverse energy is required to be greater than 120 and 150 for the two HM triggers. 
The number of reconstructed tracks by the high-level triggers~\cite{CMS:2020cmk,CMS:2016ngn} with $\abs{\eta}<2.4$, $\pt>0.4\GeVc$ is required to be greater than or equal to 120, 150, or 185. 
The events are required to contain a primary vertex, defined as the reconstructed vertex with the highest number of associated tracks, within 15\unit{cm} of 
the nominal interaction point along the beam axis and 0.2\unit{cm} in the transverse direction. 

The \pPb data are analyzed in several charged particle multiplicity classes defined according to the number of selected 
tracks (\noff) with $\abs{\eta}<2.4$ and $\pt>0.4\GeVc$, that
originate from the primary vertex, and that satisfy the high-purity criteria of Ref.~\cite{CMS:2014pgm}.
For each \noff range with at least one reconstructed \PGL or \PAGL candidate (described later), 
Table \ref{tab:ntrk-map} presents the fraction of MB triggered events found in this range, 
the average \noff value for this range $\langle\noff\rangle$, and the average number of 
tracks after correcting for track reconstruction inefficiencies and misidentification rates \ncorravg. 
With negligible statistical uncertainties, the common systematic uncertainty for the \ncorravg values is 2.4\%, independent of track multiplicity, 
resulting from the uncertainty in the tracking efficiency 
estimate based on simulated events~\cite{tracking}.
Events with $\noff<3$ are not included because of low trigger efficiency,  
while events in the multiplicity region of $\noff>250$ are not included to avoid effects 
of multiple interactions in a single event.
\begin{table}[htbp]
  \centering
  \topcaption{For each \noff range where at least one reconstructed \PGL or \PAGL candidate is found, 
  the fraction of MB triggered events found in this range, the average \noff value for this range $\langle\noff\rangle$, and the average number of tracks after correcting for track reconstruction inefficiencies and misidentification rates \ncorravg are reported.
  }
  \begin{scotch}{lccc}
    \noff & Fraction & $\langle\noff\rangle$ & \ncorravg \\
    \hline
    $[3, 60)$ & $0.66$ & $40.0$ & $48.5 \pm 1.2$\\
    $[60, 120)$ & $0.28$ & $86.7$ & $105.3 \pm 2.5$\\
    $[120, 150)$ & $0.04$ & $132.7$ & $161.2 \pm 3.9$\\
    $[150, 185)$ & $0.01$ & $163.6$ & $198.7 \pm 4.8$\\
    $[185, 250)$ & $3 \times 10^{-3}$ & $203.3$ & $246.9 \pm 5.9$\\
  \end{scotch}
\label{tab:ntrk-map}
\end{table}

The reconstruction and selection procedures for strange hadron candidates in \pPb  collisions are
identical to those in Refs. \cite{Khachatryan:2014jra,Khachatryan:2016yru,Khachatryan:2016txc,CMS:2018loe}.
Pairs of oppositely charged particle tracks with both transverse and longitudinal impact parameter significances (magnitude normalized by its uncertainty) greater than 1 with respect to the primary vertex
are selected to determine
if they point to a common secondary vertex resulting from the decay of a \PGL  candidate.
The reconstruction efficiency for tracks with low momenta and large impact parameters
is increased by using all tracks that pass the loose selection criteria \cite{CMS:2014pgm}.
When reconstructing the \PGL  (\PAGL), the two tracks are assumed to be $\Pgpm\Pp\ (\Pgpp\Pap)$. 
To remove \PKzS\ candidates misidentified as \PGL/\PAGL  particles, the two particles are assumed to be pions and the decaying particle is rejected 
if its invariant mass is less than 20\MeV away from the Particle Data Group value of the \PKzS\ mass~\cite{ParticleDataGroup:2024cfk}.

Several topological selections are applied to reduce the combinatorial background.
The selection is optimized in order to maximize the expected statistical significance of the signal. 
Strange hadron candidates are selected with the requirements that the normalized
$\chi^2$ of the Kalman Vertex Fitter \cite{CMS:2014pgm} for their decay vertex is less than 7. The three-dimensional distance (normalized by its uncertainty)
between the primary and secondary vertices is required to exceed 5. The pointing angle ($\theta^{\mathrm{point}}$), defined as the angle between the line segment
connecting the primary and secondary decay vertices and the momentum vector of the reconstructed particle candidates in the plane
transverse to the beam direction, is required to satisfy $\cos\theta^{\mathrm{point}}>0.999$. 
These selection criteria reduce the fraction of secondary \PGL particles originating from the weak decays of \PgXm\ and \PgOm\ particles to 10\% and 1\%, respectively. 
The limited number of  detected \PgXm\ and \PgOm\ particles does not allow for their corresponding polarizations to be determined, preventing a direct correction due to these secondary \PGL particles. 

As the spin of a particle is difficult to measure directly in heavy ion collisions, the parity-violating weak decays of $\PGL \to\Pp+\Pgpm$ 
and $\PAGL \to\Pap+\Pgpp$, in which the momentum direction of the daughter proton is correlated with the spin 
of the hyperon, are used to measure the polarization \cite{Lesnik:1975my}. The angular distribution of the proton in the hyperon rest frame is given by
\begin{equation}
	\frac{\rd{N}}{\rd{\Omega}}\propto1+\alpha_{\PGL}P_{\PGL}\cos\theta^{*},
\end{equation}
where $\alpha_{\PGL}$ is the hyperon decay parameter ($\alpha_{\PGL}=0.750\pm 0.009$, $\alpha_{\PAGL}=-0.758\pm 0.010$ \cite{BESIII:2018cnd}), $P_{\PGL}$ is the 
hyperon polarization, and $\theta^{*}$ is the angle between the polarization vector and the direction of the daughter baryon momentum in the hyperon rest frame. 
To measure the polarization along the beam direction, the positive $z$ direction is used as the local polarization vector to determine $\theta^{*}$.
The polarization \Pz can be estimated by averaging $\theta^{*}$ over all hyperons detected in all events \cite{STAR:2019erd},
\begin{equation}
	\Pz=\frac{\langle\cos\theta^{*}\rangle}{\alpha_{\PGL}\langle\cos^{2}\theta^{*}\rangle}.
\end{equation}
The factor $\langle\cos^{2}\theta^{*}\rangle$, which equals $1/3$ in the case of an ideal detector, is calculated directly from the data and found to range from 0.303 to 0.351 from low- to high-multiplicity events.

The event-plane angle \Psix{n} is determined by the flow vector constructed from tracks within the range of $0.3<\pt<3.0\GeVc$ and $\abs{\eta}<2.4$. 
The decay daughters of the reconstructed \PGL  (\PAGL) candidates are excluded in the event-plane determination to avoid autocorrelations.
The two components of the flow vector are given by:
\begin{equation}
  \begin{split}
	\Qnx=Q_n\cos(n\Psix{n})=\sum_i\omegai\cos(n\phii),\\
	\Qny=Q_n\sin(n\Psix{n})=\sum_i\omegai\sin(n\phii),
  \end{split} 
\label{eq:Qn}  
\end{equation} 
where \phii are the azimuthal angles of the tracks and the corresponding weights \omegai are their transverse momentum values.
To compensate for the imperfect detector acceptance, the flow vectors are recentered, with
\begin{equation}
	\Qnx^{'}=\Qnx-\langle \Qnx \rangle, \text{and}\quad \Qny^{'}=\Qny-\langle \Qny \rangle.
\end{equation}
Then, \Psix{n} is estimated from the recentered flow vector components,
\begin{equation}
    \quad\Psix{n}=\frac{1}{n}\arctan\left(\frac{\Qny^{'}}{\Qnx^{'}}\right).
\end{equation}
A standard flattening technique is then used to remove the residual nonuniformities in the event-plane angular distribution \cite{Barrette:1997pt}. 
The second-order Fourier sine coefficient, \Pzs, can be measured directly, with 
\begin{equation}
  \langle \Pz\sin[2(\phi-\Psix{2})] \rangle \\
   =\frac{\langle\cos\theta^{*}\sin[2(\phi-\Psix{2})]\rangle}{\alpha_{\PGL}\mathrm{Res}(\Psix{2})\langle\cos^{2}\theta^{*}\rangle},
\end{equation}
where $\mathrm{Res}(\Psix{2})$ is the event-plane resolution due to the finite number of final-state particles. This correction is estimated using a three-subevent method \cite{Poskanzer:1998yz},
\begin{equation}
  \mathrm{Res}(\Psix{2})=\sqrt{\frac{\langle\cos[2(\Psix{2}-\Psix{2}^{\mathrm{HF^{+}}})]\rangle \langle\cos[2(\Psix{2}-\Psix{2}^{\mathrm{HF^{-}}})]\rangle}{\langle\cos[2(\Psix{2}^{\mathrm{HF^{+}}}-\Psix{2}^{\mathrm{HF^{-}}})]\rangle}}.
\end{equation} 
The positive $\eta$ side HF and negative $\eta$ side HF are used to determine $\Psix{2}^{\mathrm{HF^{+}}}$ and $\Psix{2}^{\mathrm{HF^{-}}}$, 
where \phii and \omegai in Eq. (\ref{eq:Qn}) are the azimuthal angles of the calorimeter towers and the transverse energy deposited, respectively. 
The $\mathrm{Res}(\Psix{2})$ values range from 0.415 (low multiplicity) to 0.684 (high multiplicity). 

\begin{figure}[thbp]
\centering
\includegraphics[width=0.49\textwidth]{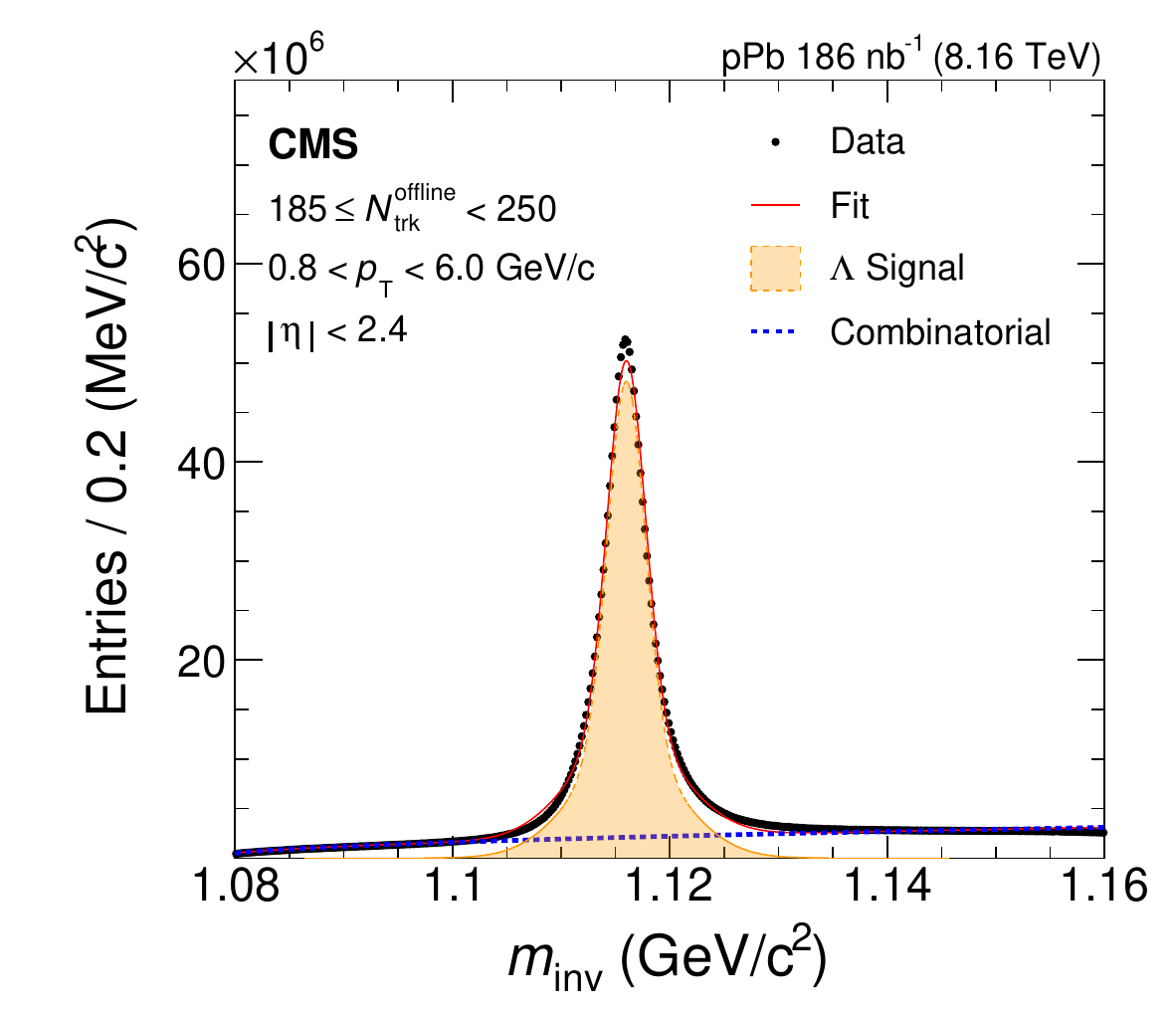}
\includegraphics[width=0.49\textwidth]{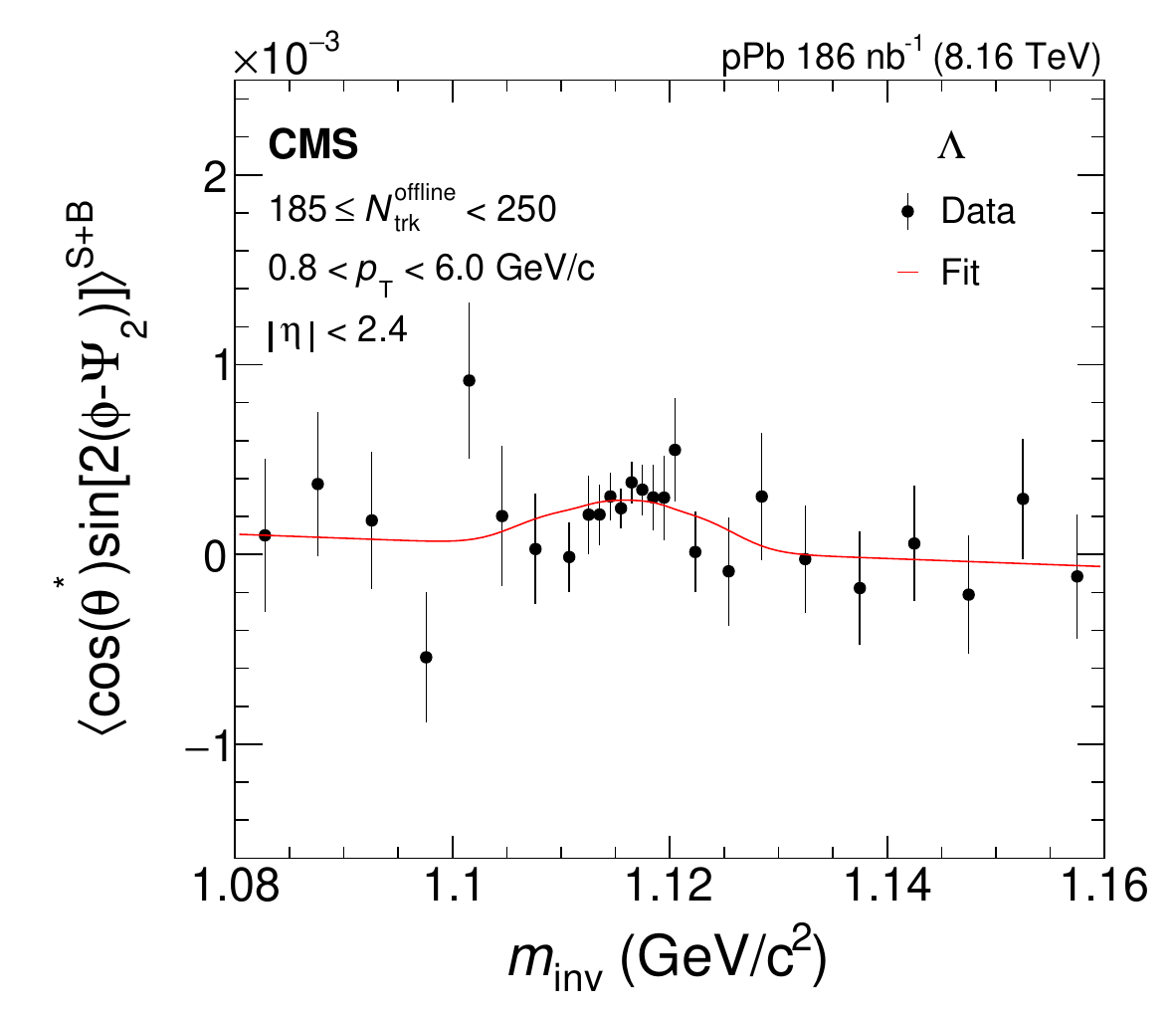}
\caption{Example of a simultaneous fit to the mass spectrum (\cmsleft) and $\langle \cos\theta^{*}\sin[2(\phi-\Psix{2})] \rangle^{S+B}$ (\cmsright) for the multiplicity range $185\leq\noff< 250$. The mass spectrum is fitted by a double Gaussian together with an additional term that is needed to account for the combinatorial background, $A_{1}q^{1/2} + A_{2}q^{3/2}$, where $q = m_{\text{inv}} - (m_{\PGp}+m_{\Pp})$. The vertical bars show the statistical uncertainties.}
\label{fig:Pz_inv_fit}
\end{figure}

To extract the polarization of the hyperon signal, a simultaneous fit to the invariant mass ($m_{\text{inv}}$) spectrum of hyperon candidates and their $\langle \Pz\sin[2(\phi-\Psix{2})]\rangle$ values as a function of $m_{\text{inv}}$ is performed. 
The mass spectrum fit function is composed of the sum of two Gaussian functions with the same mean but different widths for the hyperon signal $S(m_{\text{inv}})$ and a function $A_{1}q^{1/2} + A_{2}q^{3/2}$, where $A_{1}$ and $A_{2}$ are the fitted parameters and $q = m_{\text{inv}} - (m_{\pi}+m_{\mathrm{p}})$, with $m_{\pi}$ ($m_{\mathrm{p}}$) the pion (proton) mass, to model the combinatorial background $B(m_{\text{inv}})$. 
The $\langle \Pz \sin[2(\phi-\Psix{2})]\rangle$ distribution is fitted with 
\ifthenelse{\boolean{cms@external}}
{
  \begin{multline}
 	\langle\cos\theta^{*}\sin[2(\phi-\Psix{2})]\rangle^{S+B}\\
  =\alpha(m_{\text{inv}})\langle\cos\theta^{*}\sin[2(\phi-\Psix{2})]\rangle^{S}\\
  +[1-\alpha(m_{\text{inv}})]\langle\cos\theta^{*}\sin[2(\phi-\Psix{2})]\rangle^{B}.
  \end{multline}
}
{
  \begin{multline}
    \langle\cos\theta^{*}\sin[2(\phi-\Psix{2})]\rangle^{S+B}=\alpha(m_{\text{inv}})\langle\cos\theta^{*}\sin[2(\phi-\Psix{2})]\rangle^{S}\\
   +[1-\alpha(m_{\text{inv}})]\langle\cos\theta^{*}\sin[2(\phi-\Psix{2})]\rangle^{B}.
   \end{multline}
}
Here $\langle\cos\theta^{*}\sin[2(\phi-\Psix{2})]\rangle^{B}$ for the background candidates is modeled as a linear function of the invariant mass and $\alpha(m_{\text{inv}})$ is the hyperon signal fraction. 
Figure \ref{fig:Pz_inv_fit} shows an example of a simultaneous fit to the mass spectrum and $\langle\cos\theta^{*}\sin[2(\phi-\Psix{2})]\rangle^{S+B}$ for the multiplicity range $185\leq\noff<250$.

The sources of systematic uncertainty in the \Pzs measurement include the variation of the background probability density function (PDF) 
for the invariant mass and, 
for the $\langle\cos\theta^{*}\sin[2(\phi-\Psix{2})]\rangle$ value of background \PGL candidates, 
the vertex position, 
variation of the results when the proton and Pb beam directions are switched, and the uncertainty on the hyperon decay parameter \cite{BESIII:2018cnd}.
Systematic effects related to background invariant mass PDF are investigated by changing the PDF to a third-order polynomial, an exponential, 
and a second- or third-order Chebyshev polynomial functions. 
Based on these studies, a 3\% uncertainty for the \Pzs results is assigned. 
The systematic uncertainties from the background $\langle\cos\theta^{*}\sin[2(\phi-\Psix{2})]\rangle^{B}$ functional form are evaluated by changing from a linear function to a constant background,  including the possibility of this background being negligible. 
For $\noff<120$, the resulting uncertainty is 3\% (7\%) for \PGL (\PAGL), and 5\% for \PGL + \PAGL results. This uncertainty is found to be negligible for the higher-multiplicity ranges.
To evaluate the uncertainties from the vertex-dependent detector acceptance, the \Pzs are extracted from events with $\abs{v_z}<3\cm$ and $3<\abs{v_z}<15\cm$. 
A variation of 8\% (3\%) is found for the \PGL/\PAGL (\PGL + \PAGL) results, which is quoted as a systematic uncertainty. 
The beam directions were reversed during the 2016 \pPb collision data taking, and a systematic uncertainty of 10\% is found by comparing results from the two beam directions. 
The hyperon decay parameter measurements have their own uncertainties, which lead to global systematic uncertainties of 1.2\% (1.3\%) for \PGL (\PAGL) results and 1.3\% for the \PGL + \PAGL results. 
To verify the procedure for extracting the signal \Pzs value, a study using {\textsc{epos lhc}} \cite{Pierog:2013ria} \pPb events was performed, with the extracted values found to be consistent with the generator-level values.
To confirm the absence of other unexpected detector effects, \Pzs for \PKzS\ was also measured using the same analysis procedure as for \PGL (\PAGL). 
The results are consistent with zero, as expected for a spin-0 particle, and are shown in \refsupplemental. 

\begin{figure}[htbp]
  \centering
  \includegraphics[width=0.49\textwidth]{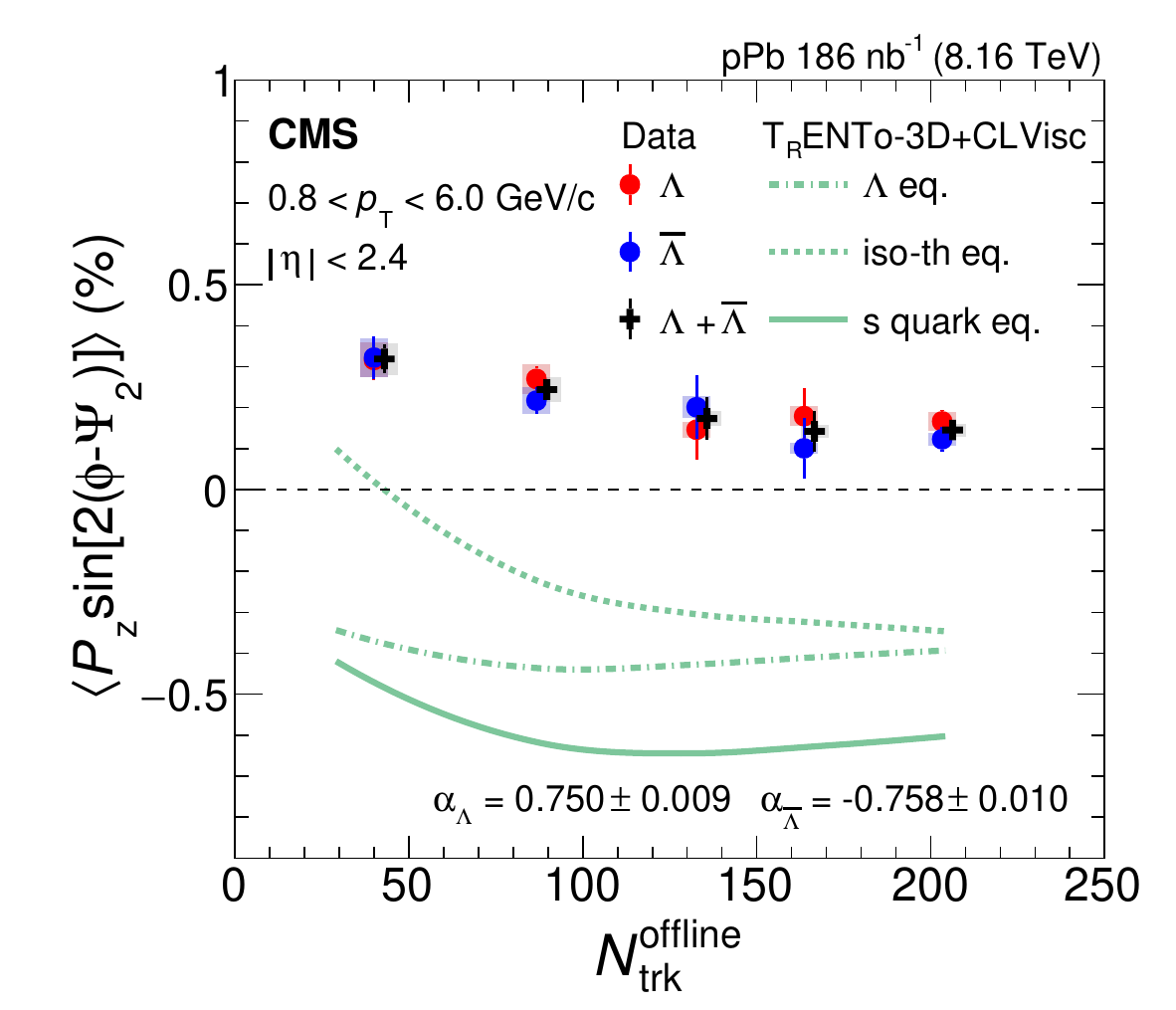}
  \includegraphics[width=0.49\textwidth]{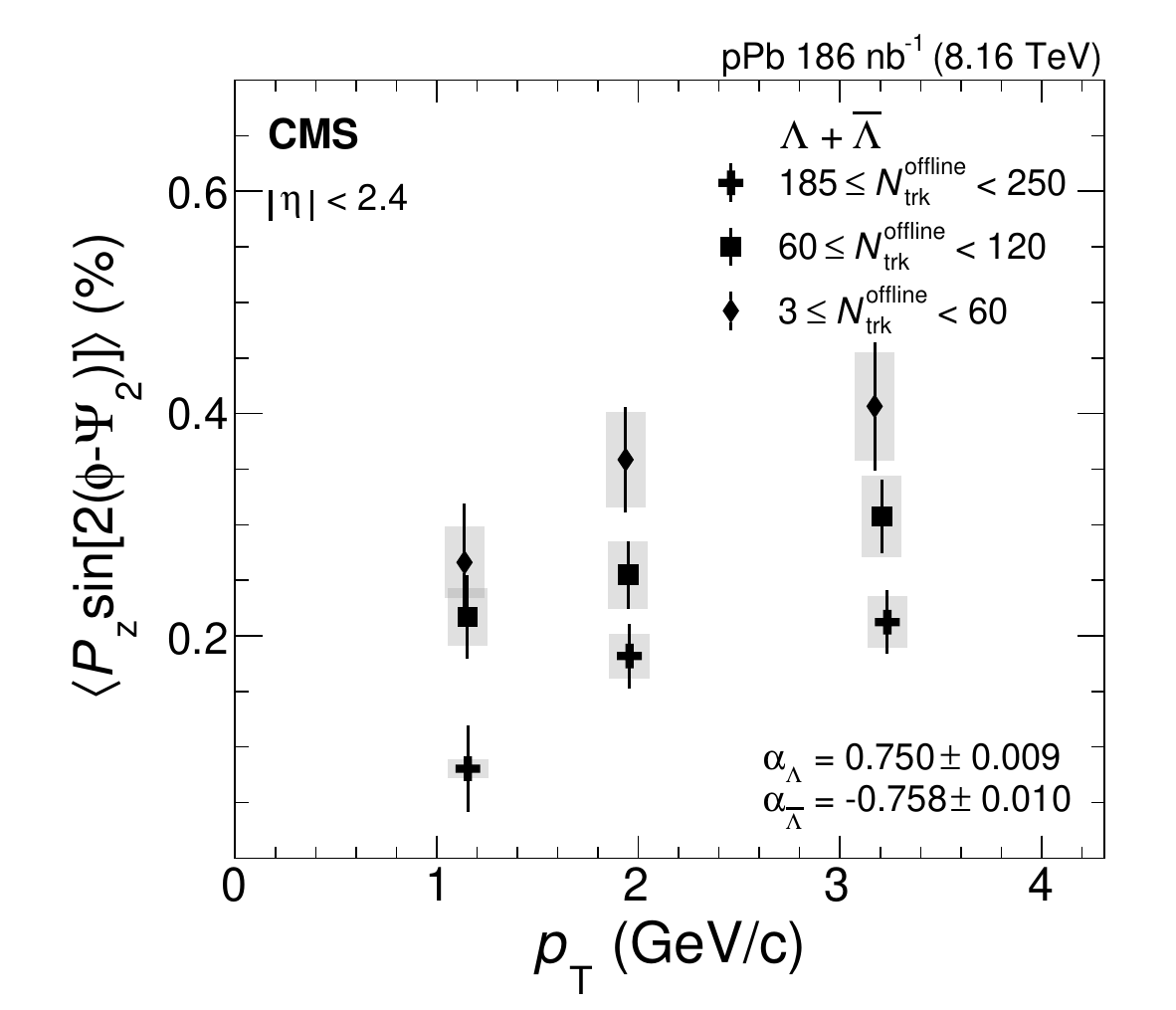}
  \caption{\cmsLeft: The second-order Fourier sine coefficients of \PGL, \PAGL  and \PGL + \PAGL polarizations along the beam direction as functions of \noff\ in \pPb collisions at $\sqrtsNN = 8.16\TeV$. Results from hydrodynamic calculations \cite{Yi:2024kwu} are shown as solid and dashed lines. The \noff\ values of \PGL + \PAGL results are shifted horizontally for better visibility. \cmsRight: The second-order Fourier sine coefficients \PGL + \PAGL polarization along the beam direction as functions of \pt for $3 \leq \noff < 60$, $60 \leq \noff < 120$, and $185 \leq \noff < 250$ in \pPb collisions at $\sqrtsNN = 8.16\TeV$. Vertical bars represent statistical uncertainties, while shaded areas show systematic uncertainties.}
  \label{fig:Pz_Ntrk}
\end{figure}

Figure \ref{fig:Pz_Ntrk} (\cmsleft) shows the \noff\ dependence of \Pzs in \pPb collisions at $\sqrtsNN = 8.16\TeV$ for the \PGL and \PAGL hyperons, as well as for the combined \PGL + \PAGL results. 
Positive \Pzs values are observed, with a significance of nonzero polarization exceeding 5 standard deviations for three of the five \noff\ bins.
The \Pzs values for \PGL  and \PAGL  particles are consistent with each other within the uncertainties and found to decrease as a function of multiplicity. 
A similar trend was observed in \AonA collisions, where the \Pzs values are found to decrease as collisions become more central (\ie, with increasing multiplicity)~\cite{STAR:2019erd,ALICE:2021pzu,STAR:2023eck}. 
This trend deviates from the usual behavior of other collective flow observables, which typically diminish at low multiplicities, consistent with hydrodynamic models predicting a progressively shorter QGP lifetime \cite{Nagle:2018nvi, Dusling:2015gta}. 
In particular, the \Pzs values show a different trend than the elliptic flow, which is decreasing toward low multiplicity \cite{CMS:2013jlh}. 
This discrepancy indicates that, in addition to the potential polarization induced by hydrodynamical expansion, there might be other mechanisms contributing to the measured \Pzs.
To ensure that the measured \Pzs values are not affected by autocorrelations, the measurements have been repeated using the HF calorimeter on the Pb-going side of the CMS detector, with an acceptance range of $3.0<\abs{\eta}<5.0$, to determine the event-plane angle. 
The results are found to be consistent with those obtained using the tracker for the event-plane angle determination. The comparison is shown in \refsupplemental. 
The \Pzs results for the \PGL + \PAGL  particles are further studied as a function of \pt\ for $3\leq\noff<60$, $60\leq\noff<120$, and $185 \leq\noff<250$, which are shown in Fig. \ref{fig:Pz_Ntrk} (\cmsright). 
For \pt above $\sim$1.5\GeVc, the significance exceeds 5 standard deviations for all of the \noff\ bins. 
An increasing trend as a function of \pt\ is observed for all measured multiplicity ranges. 

The \Pzs values estimated from a hydrodynamic model including the thermal shear and thermal vorticity contributions~\cite{Yi:2024kwu} are shown in Fig.~\ref{fig:Pz_Ntrk} for comparison. 
The 3+1D CLVisc model~\cite{Pang:2018zzo,Wu:2021fjf} with the parameterized T\raisebox{-0.5ex}{R}ENTo initial conditions~\cite{Moreland:2018gsh} is used for the hyperon polarization calculation.
The model is tuned to describe the charged hadron multiplicity and the transverse-momentum-dependent elliptic flow of \PGL hyperons in \pPb collisions at $\sqrtsNN = 8.16\TeV$. 
Three different scenarios commonly used to model spin polarization in relativistic \AonA collisions~\cite{Becattini:2017gcx,Xia:2018tes,Fu:2021pok,Becattini:2021iol,Alzhrani:2022dpi,Florkowski:2021xvy} are studied. 
In the \PQs quark equilibrium scenario, the spin of \PGL hyperons is assumed to be carried by the constituent \PQs quark based on the coalescence model~\cite{Liang:2004ph}.
In the \PGL equilibrium scenario, the \PGL hyperon polarization is calculated at the freeze-out, where the mass of the spin carrier is the hyperon mass~\cite{Becattini:2013fla}. 
In addition to the \PGL equilibrium scenario, the isothermal equilibrium scenario assumes a constant temperature of the system at the freeze-out \cite{Becattini:2021suc}. 
The thermal shear and thermal vorticity contributions to \Pzs in these calculations are found to be positive and negative, respectively. 
While the three approaches describe the positive \Pzs values observed in \AonA collisions quantitatively or qualitatively~\cite{ALICE:2021pzu,Alzhrani:2022dpi,Becattini:2021iol}, 
the calculations in \pPb collisions lead to negative \Pzs values which disagree with the measured results. 

The new \pPb data pose a challenge to the current theoretical implementation of 
spin polarization in heavy ion collisions and underscore the importance of 
investigating other physics mechanisms, beyond QCD vorticity effects, 
which might also explain the observed hyperon polarization in proton-nucleus (\pA) and \AonA collisions.
For example, the general polarization of \PGL hyperons produced in higher-energy collisions of unpolarized hadrons 
has been a major challenge for QCD theoretical interpretations for several decades~\cite{Bunce:1976yb,Heller:1978ty}.
Recently, the transverse polarization of \PGL hyperons has been measured in unpolarized $\Pep\Pem$ collisions by the Belle Collaboration \cite{Belle:2018ttu},
with the effects of polarizing fragmentation functions (PFFs) proposed to describe the measured polarization~\cite{Anselmino:2019cqd,DAlesio:2020wjq,Li:2020oto,Gamberg:2021iat}. 
Further investigations are needed to understand how PFFs and their potential dependence on parton flavor~\cite{Gamberg:2021iat} contribute to \PGL hyperon polarization in hadronic collisions at LHC energies.

In summary, the first measurement of the second-order sine Fourier coefficients \Pzs of hyperon polarization along the beam direction in proton-lead collisions is presented. 
The data were obtained using the CMS detector at a nucleon-nucleon center of mass energy of  8.16\TeV. 
Significant positive \Pzs values are observed for \PGL  and \PAGL  particles as a function of charged particle multiplicity and transverse momentum. 
The observed signal exhibits behavior similar to that seen in nucleus-nucleus collisions, with \Pzs values increasing with rising transverse momentum and decreasing charged particle multiplicity.
Hydrodynamic calculations with various polarization scenarios, which describe the results in nucleus-nucleus collisions, fail to describe the positive sign of the proton-lead results. These results pose a challenge to the current theoretical interpretation of spin polarization in heavy ion collisions.

\begin{acknowledgments}
  We congratulate our colleagues in the CERN accelerator departments for the excellent performance of the LHC and thank the technical and administrative staffs at CERN and at other CMS institutes for their contributions to the success of the CMS effort. In addition, we gratefully acknowledge the computing centers and personnel of the Worldwide LHC Computing Grid and other centers for delivering so effectively the computing infrastructure essential to our analyses. Finally, we acknowledge the enduring support for the construction and operation of the LHC, the CMS detector, and the supporting computing infrastructure provided by the following funding agencies: SC (Armenia), BMBWF and FWF (Austria); FNRS and FWO (Belgium); CNPq, CAPES, FAPERJ, FAPERGS, and FAPESP (Brazil); MES and BNSF (Bulgaria); CERN; CAS, MoST, and NSFC (China); MINCIENCIAS (Colombia); MSES and CSF (Croatia); RIF (Cyprus); SENESCYT (Ecuador); ERC PRG, RVTT3 and MoER TK202 (Estonia); Academy of Finland, MEC, and HIP (Finland); CEA and CNRS/IN2P3 (France); SRNSF (Georgia); BMBF, DFG, and HGF (Germany); GSRI (Greece); NKFIH (Hungary); DAE and DST (India); IPM (Iran); SFI (Ireland); INFN (Italy); MSIP and NRF (Republic of Korea); MES (Latvia); LMTLT (Lithuania); MOE and UM (Malaysia); BUAP, CINVESTAV, CONACYT, LNS, SEP, and UASLP-FAI (Mexico); MOS (Montenegro); MBIE (New Zealand); PAEC (Pakistan); MES and NSC (Poland); FCT (Portugal);  MESTD (Serbia); MCIN/AEI and PCTI (Spain); MOSTR (Sri Lanka); Swiss Funding Agencies (Switzerland); MST (Taipei); MHESI and NSTDA (Thailand); TUBITAK and TENMAK (Turkey); NASU (Ukraine); STFC (United Kingdom); DOE and NSF (USA).
\end{acknowledgments}

\bibliography{auto_generated} 

\ifthenelse{\boolean{cms@external}}{}{
\clearpage
\appendix
\numberwithin{table}{section}
\numberwithin{figure}{section}
\section{Supplemental material: \texorpdfstring{\PGL}{Lambda} hyperon local polarization with forward rapidity event-plane and \texorpdfstring{\PKzS}{K} polarization}
\label{this:supplemental}
The \Pzs values for \PKzS\ particles have been measured, with $\cos\theta^{*}$ calculated using the higher \pt pion from the \PKzS\ decay.
The results, shown in Fig.~\ref{fig:Pz_Ntrk_K0S} as a function of \noff for \pPb collisions at $\sqrtsNN = 8.16\TeV$,
are consistent with zero,  as expected for a spin-0 particle.
\begin{figure}[tbh!]
    \centering
    \includegraphics[width=0.49\textwidth]{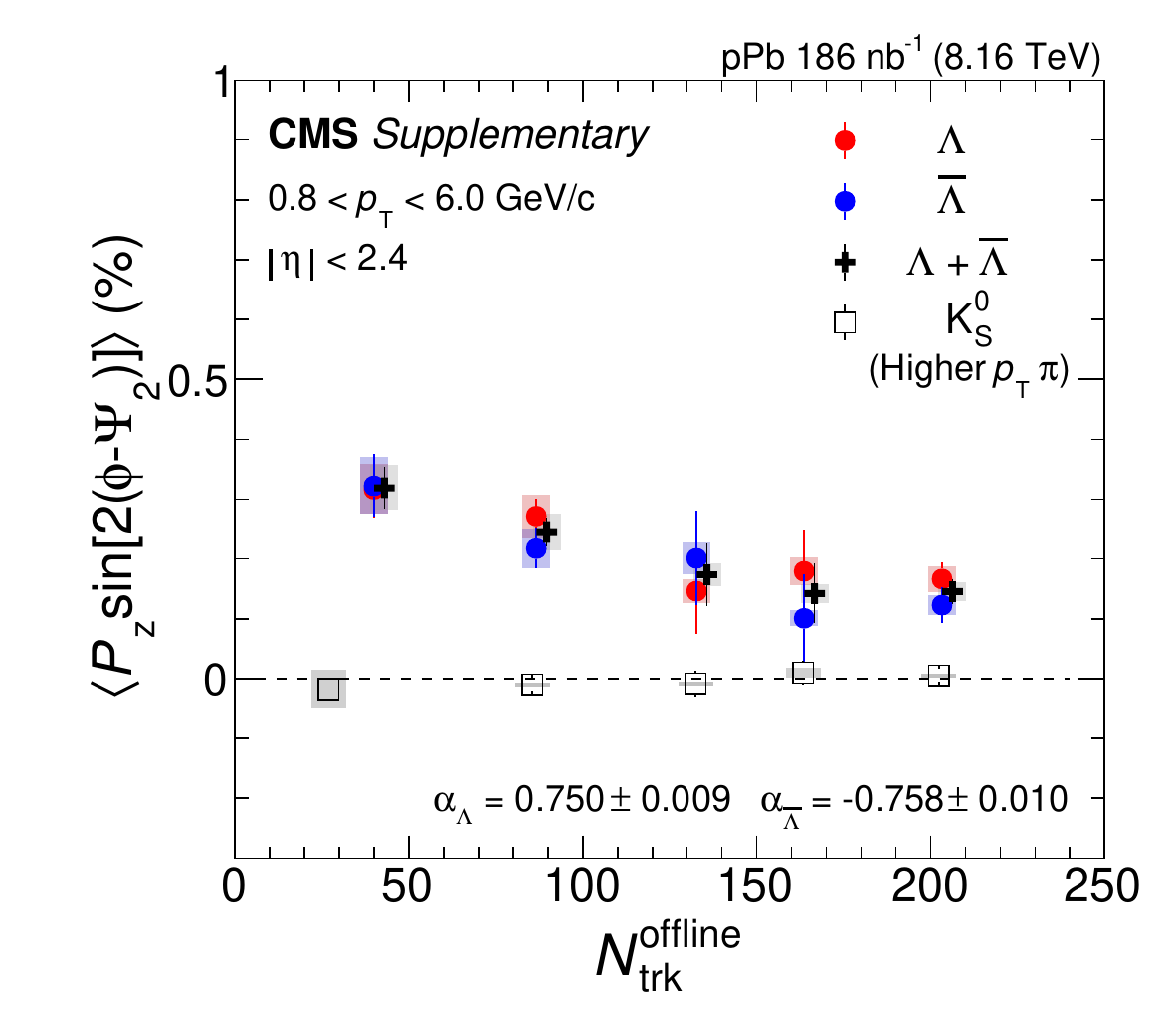}
    \caption{The second-order Fourier sine coefficients of \PKzS with $\cos\theta^{*}$ calculated from the higher \pt $\pi$ from \PKzS\ decay, \PGL, \PAGL and \PGL + \PAGL polarization along the beam direction as functions of \noff\ in \pPb collisions at $\sqrtsNN = 8.16\TeV$. Vertical bars show statistical uncertainties. Shaded areas show systematic uncertainties. The \noff\ values of \PGL + \PAGL results are shifted for better visibility.}
    \label{fig:Pz_Ntrk_K0S}
\end{figure}

To ensure that the measured \Pzs values for the \PGL hyperon is not affected by short range correlations,
the measurements have been repeated using event planes reconstructed with the HF detectors.  These detectors have a significantly different $\eta$ range ($2.9 < \abs{\eta} < 5.2$) than the observed \PGL hyperons.
The HF-based event planes are reconstructed using only one side of the HF, corresponding to the Pb beam direction and denoted as $\Psix{2}^{\mathrm{HF_{Pb-going}}}$, to achieve improved event plane resolution.
The results, shown in Figs.~\ref{fig:Pz_Ntrk_HF_La_no_pgoing}--\ref{fig:Pz_Ntrk_HF_LAL_no_pgoing}, are found to be consistent with those measured using the event planes reconstructed with mid-rapidity tracks.
\begin{figure}[tbh!]
\centering
\includegraphics[width=0.49\textwidth]{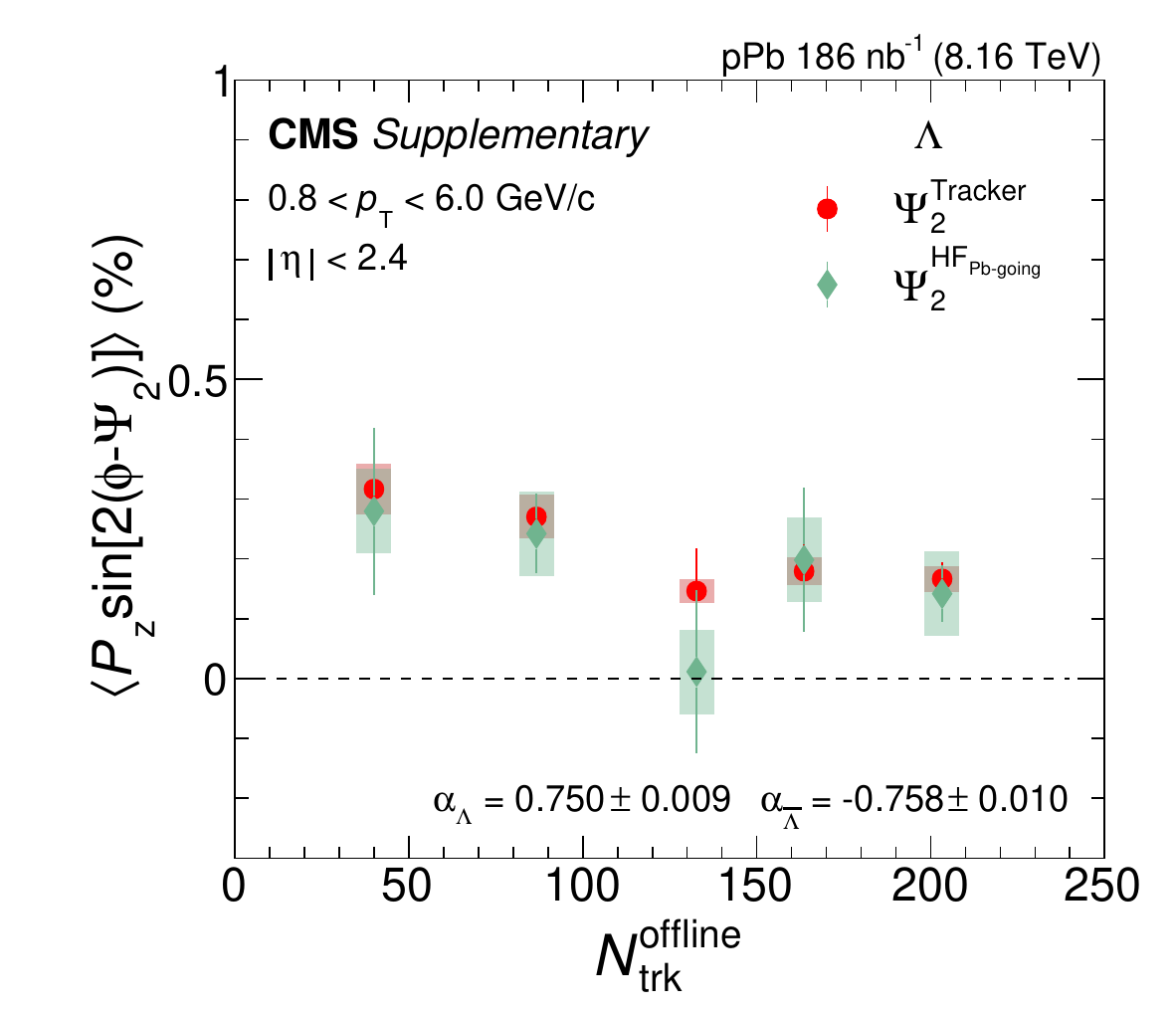}
\caption{The second-order Fourier sine coefficients of \PGL polarization along the beam direction as functions of \noff\ in \pPb collisions at $\sqrtsNN = 8.16\TeV$ extracted with tracker and HF based event planes. Vertical bars show statistical uncertainties. Shaded areas show systematic uncertainties.}
\label{fig:Pz_Ntrk_HF_La_no_pgoing}
\end{figure}
\begin{figure}[tbh!]
\centering
\includegraphics[width=0.49\textwidth]{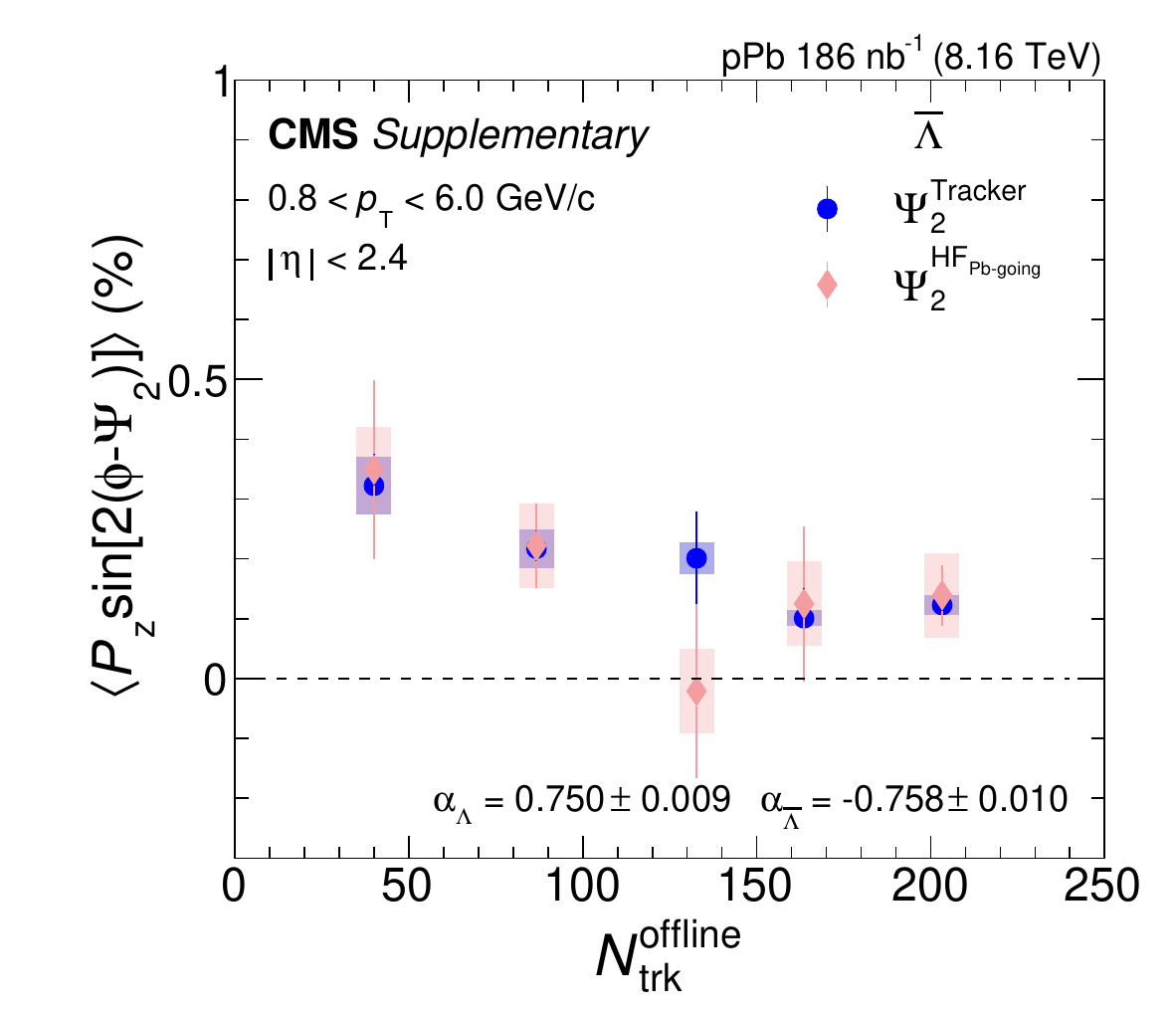}
\caption{The second-order Fourier sine coefficients of \PAGL polarization along the beam direction as functions of \noff\ in \pPb collisions at $\sqrtsNN = 8.16\TeV$ extracted with tracker and HF based event planes. Vertical bars show statistical uncertainties. Shaded areas show systematic uncertainties.}
\label{fig:Pz_Ntrk_HF_ALa_no_pgoing}
\end{figure}
\begin{figure}[tbh!]
\centering
\includegraphics[width=0.49\textwidth]{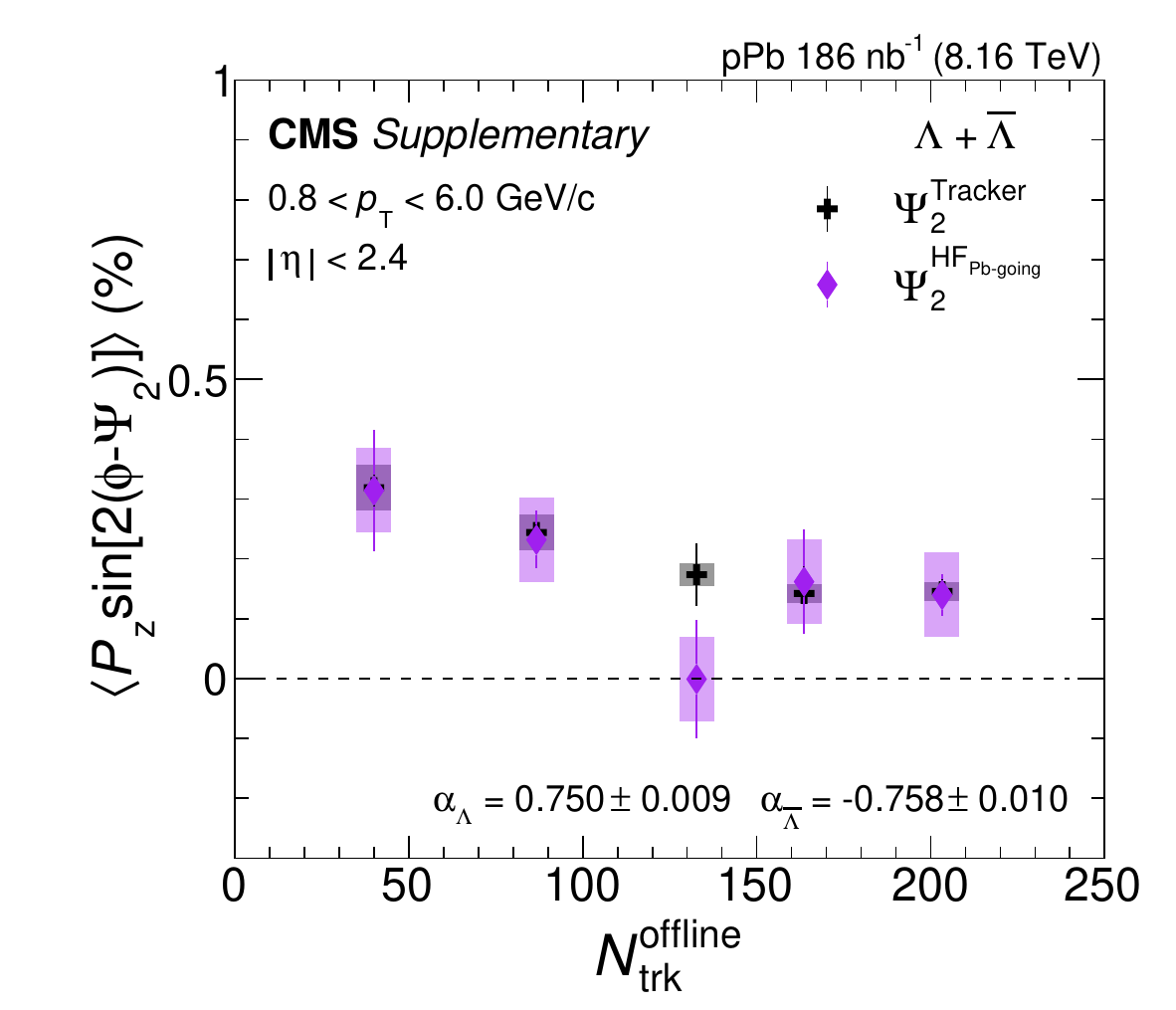}
\caption{The second-order Fourier sine coefficients of \PGL + \PAGL polarization along the beam direction as functions of \noff\ in \pPb collisions at $\sqrtsNN = 8.16\TeV$ extracted with tracker and HF based event planes. Vertical bars show statistical uncertainties. Shaded areas show systematic uncertainties.}
\label{fig:Pz_Ntrk_HF_LAL_no_pgoing}
\end{figure}

}\cleardoublepage \section{The CMS Collaboration \label{app:collab}}\begin{sloppypar}\hyphenpenalty=5000\widowpenalty=500\clubpenalty=5000
\cmsinstitute{Yerevan Physics Institute, Yerevan, Armenia}
{\tolerance=6000
A.~Hayrapetyan, A.~Tumasyan\cmsAuthorMark{1}\cmsorcid{0009-0000-0684-6742}
\par}
\cmsinstitute{Institut f\"{u}r Hochenergiephysik, Vienna, Austria}
{\tolerance=6000
W.~Adam\cmsorcid{0000-0001-9099-4341}, J.W.~Andrejkovic, L.~Benato\cmsorcid{0000-0001-5135-7489}, T.~Bergauer\cmsorcid{0000-0002-5786-0293}, S.~Chatterjee\cmsorcid{0000-0003-2660-0349}, K.~Damanakis\cmsorcid{0000-0001-5389-2872}, M.~Dragicevic\cmsorcid{0000-0003-1967-6783}, P.S.~Hussain\cmsorcid{0000-0002-4825-5278}, M.~Jeitler\cmsAuthorMark{2}\cmsorcid{0000-0002-5141-9560}, N.~Krammer\cmsorcid{0000-0002-0548-0985}, A.~Li\cmsorcid{0000-0002-4547-116X}, D.~Liko\cmsorcid{0000-0002-3380-473X}, I.~Mikulec\cmsorcid{0000-0003-0385-2746}, J.~Schieck\cmsAuthorMark{2}\cmsorcid{0000-0002-1058-8093}, R.~Sch\"{o}fbeck\cmsAuthorMark{2}\cmsorcid{0000-0002-2332-8784}, D.~Schwarz\cmsorcid{0000-0002-3821-7331}, M.~Sonawane\cmsorcid{0000-0003-0510-7010}, W.~Waltenberger\cmsorcid{0000-0002-6215-7228}, C.-E.~Wulz\cmsAuthorMark{2}\cmsorcid{0000-0001-9226-5812}
% ...
\par}
\vskip\cmsinstskip
\dag:~Deceased\\
$^{1}$Also at Yerevan State University, Yerevan, Armenia\\
$^{2}$Also at TU Wien, Vienna, Austria\\
% ...
\end{sloppypar}
\end{document}